# Housing Price Prediction Model Selection Based on Lorenz and Concentration Curves: Empirical Evidence from Tehran Housing Market

Mohammad Mirbagherijam[1]
Assistant Professor in Economics / Shahrood University of Technology
**Email:** m.mirbagherijam@shahroodut.ac.ir

**Abstract**
This study contributes a house price prediction model selection in Tehran city based on the area between Lorenz and concentration curves of predicted price by using 206,556 observed transaction data over the period from March 21, 2018 to February 19, 2021. Several different methods such as "generalized linear models" (GLM) and "recursive partitioning and regression trees" (RPART), "random forests" (RF) regression models and "neural network" (NN) models were examined for house price prediction. We used 90% of all data samples which were chosen randomly to estimate the parameters of pricing models and 10% of remaining datasets to test the accuracy of prediction. Results showed that the area between the LC and CC curves (which are known as ABC criterion) of real and predicted price in the test data sample by random forest regression model was less than by other models under study. The comparison of the calculated ABC criteria lead us to conclude that the nonlinear regression models such as RF regression models give an accurate prediction of house price in Tehran city.

## 1. Introduction

House pricing and house price prediction are two related major topics of interest to researchers in the field of housing economics. It could be due to the benefit of accurate house price prediction and proper house pricing for participant of house market, especially for the tenants and landlords, buyers and sellers, real estate agents and appraisers, government and mortgage lenders, as is well known and commonly observed in the literature (Oladunni et al., 2017).

---

[1] Corresponding Author;

Department of Industrial Engineering and Management, Shahrood University of Technology, Iran.
P.O. Box 3619995161; Tel.: +982332392204-9 ; Mobile.: +98 912 758 5015;
Homepage: https://shahroodut.ac.ir/en/as/?id=S860

Accuracy of prediction which is measured by an association of pairs of real and predicted or expected prices is the primary concern of house price forecasting. This is owing to decision making by market participants based on inaccurate forecasts that might cause them losses.

In general, the nature of housing assets has made it to some extent difficult to model and predict its price accurately. Because a house is an immovable multi-purpose asset transacted for residence and investment purpose and, therefore, a large number of features and variables affect the house price. Determinants of house price vary from changeable factors such as macroeconomic variables to fixed attributes such as region and skeleton type of a building. Hong, Choi, and Kim (2020) classified the house price factors into four categories, i.e. structural attributes[1], neighborhood attributes[2], locational attributes[3] (such accessibility to nearby facilities), and macroeconomic variable factors. Identifying the main determinant of house price and quantifying the effect of each variable or a group of them on the house price are two crucial things in modeling and forecasting the house price accurately. However, the relative importance of relevance house price factors is not the same and that might have changed over time as that might have changed from place to place. This is owing to several causes including variation of economic condition, changes in government policies, change of construction technology of building, and change of spatial house price determinants due to the lack of uniform development of urban, and rising the density of urban. This is why modeling and accurate prediction of house price remained as an open object in the literature.

To date, various methods have been developed to model and predict the house price. Hedonic pricing models (for example Malpezzi, 2008; Hu et al., 2013; Oladunni et al., 2017), principal component analysis (PCA), partial least squares (PLS), and sparse PLS (SPLS) approaches (Bork and Møller, 2018), structural time series models (STSM) (Mousavi and Doroodian, 2016), random forest (RF) method (Antipov and Pokryshevskaya, 2012; Čeh et al., 2018; Hong et al., 2020), artificial neural network (ANN) methods (Selim, 2009; Chiarazzo et al., 2014; Fachrurrazi et al., 2017), fuzzy logic (FL) models (Kuşan, Aytekin, and Özdemir, 2010; Sarip et al., 2016) and machine learning (ML) algorithms (Park and Kwon Bae, 2015; Trawinski et al., 2017; Banerjee and Dutta, 2018; Pérez-Rave et al., 2019; Jarosz et al., 2020; Truong et al., 2020) are typical examples of methods used in house price forecasting and appraising found in the literature. However, in some studies, measuring and improving the accuracy of prediction have not been addressed.

Regarding the variety of methods and models in the prediction of house price, measurement criteria of forecast accuracy and performance of models are also different in the literature. MSE or root MSE (RMSE), mean absolute error (MAE), and mean absolute percentage error (MAPE) are the most common metrics used in many studies. Therefore, these metrics have been exemplified as top three performance metrics in a report (Botchkarev, 2019).

---

[1]. Construction year, region, floor level, and type of heating system show the structural attributes.

[2]. Apartment brand, available units in the building, number of buildings in the apartment complex, parking lot, floor region ratio, building coverage ratio, and the top/lowest floor of the building are known as neighborhood attributes.

[3]. Latitude, longitude, and accessibility to nearby facilities related to the geographical position of house and categorized as the locational attributes.

Performance metric refers to a "logical and mathematical construct designed to measure how close the actual results are from what has been expected or predicted." To better understand the structure and properties of accuracy metrics, Botchkarev (2019) classified them into four categories, i.e. primary, extended, composite, and hybrid sets of metrics. Furthermore, he highlighted the differences of primary metrics by method of determining point distance, method of normalization, and method of aggregation of point distances over a data set of actual data and predicted results.

To improve the house price prediction, several practical solutions can be used. One might express them under the following headings: (1) Developing the house price forecasting models as reported previously developed from simple form, such hedonic pricing model, to novel form, such as machine learning models. (2) Employing a combination of methods to model and forecast. By way of illustration, Atrianfar et al. (2013) employed forecast combination methods including 81 effective variables, and showed how the accuracy of Tehran house price forecasting improved. Recent cases reported by Glennon et al. (2018) and Wei et al. (2020) also support forecast combination methods. (3) Including more relevant variables into the model by using large data sets. (4) Comparing the accuracy of price prediction models and choosing the more accurate model. This solution can be clearly seen in Zietz and Traian (2014) and recently in Mukhlishin et al. (2018) which set out to compare the artificial neural network with fuzzy logic methods.

Typically, the above raised concerns in improving the prediction form the basis of machine learning workflow (see Lang et al., 2019). Thereby, advances in machine learning approaches and access to big data sources will make the problem of predictive improvement much easier.

As previously stated, there are various metrics to measure the accuracy of prediction. Hence, choosing one of them as an eminently suitable metric for measuring the accuracy of forecast is difficult practically. So far, however, there has been little discussion about that. Denuit et al. (2019) showed that concentration curves (CC) and Lorenz curves (LC) provided effective tools to evaluate or to compare performance of different price prediction models. They indicated that the area between two curves CC and LC, hereafter shown by ABC, was a better indicator of the performance of a given predictor.

The main purpose of this study is to present an accurate housing price prediction model for Tehran city. It seeks to answer the question that, which of the competitive housing price prediction models are able to predict the housing price accurately. Measuring and comparing the accuracy of each housing prediction model with an eminently suitable metric such as ABC measure leads us to select more accurate prediction models properly.

There is considerable merit in taking Tehran city as the research sample. Tehran is the capital and largest populous city of Iran with about 11% of the total share of the Iranian housing market based on the resident population, and more than 43% of its resident households are tenants. However, over 20% of residential units are vacant at present. Therefore, providing accurate house price prediction models might be useful for the regulatory body to collect a proportional tax on housing prices for vacant units as it might be beneficial for other participants of the market.

The study offers some important insights into Tehran house market, and provides an important opportunity to advance the understanding of application of CC and LC in model selection metrics for accurate prediction of house price. The empirical results of research contribute to the literature of housing price prediction and model selection metrics. Several families of house price prediction models are going to be studied. This includes generalized linear models (GLM), recursive partitioning and regression trees (RPART), random forest (RF) models and neural networks (NN) models. The parameters of research models are estimated by the statistical software R. Data of housing transaction prices were collected from the website of the "Ministry of Roads and Urban Development" of Iran[1]. We used 90% of all data samples which were chosen randomly to estimate the parameters of price prediction models and 10% of the remaining data set to test the accuracy of prediction. Following Denuit et al. (2019), both concentration and Lorenz curves simultaneously used to measure the performance of the estimated models. The more accurate model for house price prediction is selected by the criterion of ABC (area between CC and LC curves) related to the set of real and predicted house price.

Due to the practical constraints, this paper cannot provide a comprehensive review of the model performance metrics. Furthermore, not all models and approaches of housing price forecasting have been investigated in the research, including STSM, PCA, PLS and SPLS approaches, fuzzy logic models, and machine learning models.

The remainder of this paper is organized as follows. Section 2 explains the model selection metric. Section 3 is divided into four subsections; the first subsection describes the research variables; the second subsection compares the relevancy of the selected variables to house price; the third subsection presents the estimation results of the models; the forth subsection evaluates the performance of the prediction models under study. Section 4 discusses findings. Finally, Section 5 concludes the paper.

**2. Model Selection Metrics**

To select the more accurate prediction model, the accuracy of several competitive house price prediction models was measured and compared by ABC criteria. ABC is detailed by Denuit et al. (2019), and refers to the area between the CC and LC.

Assume that $y_i^a$ and $y_i^p$ are the real and predicted data of house price transacted in the $i$th house transaction. Predicted price $y_i^p$, obtained by the house price prediction model π(**x**), i.e $y_i^p = \pi(x_i)$, and all house details and explanatory variables of house price were gathered in vector **x**=**x**($x_1$, $x_2$,…,$x_k$). The prediction model of π(**.**) is unknown, and we assumed that there were alternative prediction models of $\pi^1, \pi^2, ..., \pi^m$ which someone used each of them to predict the house price. To decide which model is better than another, the ABC index of each model are calculated in the following steps: First, the house prices are predicted with several different models. It is worth mentioning that before this step, the coefficients of the assumed models are estimated with the available data using R software. Then the Lorenz curve and the concentration curve of the prediction results of each model are plotted simultaneously in a graph, and the area between these two is calculated. Finally, the calculated ABC indices are compared, and the model with the lowest value of the ABC index is selected as the appropriate model. The Package of IC2 in R was used to estimate and plot the CC and LC curves. To better explain the calculation process, the definition of concentration and Lorenz curves is followed based on Denuit et al. (2019).

---

[1]. htttps://www.mrud.ir/en

**The Concentration and Lorenz Curves**

For every probability level $\alpha$, the concentration curve of the real price Y with respect to the predicted price $\pi(X)$ based on the information in the vector **x** is defined as follows[1].

$$CC[Y, \pi(X); \alpha] = \frac{E[Y|[\pi(X) \leq F_\pi^{-1}(\alpha)]]}{E[Y]} \qquad (1)$$

Where $F_\pi(t)$ is the distribution function of the predicted price ($\pi(X)$), and $F_\pi^{-1}$ is the associated with the quantile function defined as the generalized inverse of $F_\pi$, i.e. $F_\pi^{-1}(\alpha) = \inf\{t | F_\pi(t) \geq \alpha\}$ for a probability level $\alpha$. Equation 1 can be interpreted as the proportion of real price observations attributable Y to a subset of price-predicted observations at the a percentage of the lowest transactions price forecast.

The Lorenze curve LC associated with the predicted price $\pi(X)$ is as Equation 2:

$$LC[\pi(X); \alpha] = CC[\pi(X), \pi(X); \alpha] = \frac{E[\pi(X)|[\pi(X) \leq F_\pi^{-1}(\alpha)]]}{E[Y]} \qquad (2)$$

If the predicted price equals the real price, then there is no need to distinguish CC from LC. This is because if $Y = \pi(X)$, then $LC[\pi(X); \alpha] = CC[Y, \pi(X); \alpha]$.

**Properties of CC and LC**

According to Denuit et al. (2019), CC and LC curves have several certain properties. The CC and LC curves are non-decreasing (or monotonic) and convex functions. Monotonicity of CC curve satisfies $\lim_{\alpha \to 0} CC[\pi(X); \alpha] = 0$ and $\lim_{\alpha \to 1} CC[\pi(X); \alpha] = 1$.

Concentration curve is the copula of pairs $(Y, \pi(X))$. Furthermore, the area between Line 45-degree and CC measure the dependency of variables. Whenever two variables are mutually independent, the concentration curve is the line 45-degree. This line is referred

---

[1]. Assuming the samples $(y_i^a, y_i^p)$, i=1,…,n, to be independent and identically distributed, the empirical concentration curve and Lorenze curve of the real price can be estimated as follows:

$$\widehat{CC}[Y, \pi(X); \alpha] = \frac{1}{n\bar{Y}} \sum_{i | [\hat{\pi}(X_i) \leq \hat{F}_\pi^{-1}(\alpha)]} Y_i = \frac{\sum_{i | [\hat{\pi}(X_i) \leq \hat{F}_\pi^{-1}(\alpha)]} Y_i}{\sum_{i=1}^n Y_i}$$

$$\widehat{LC}[\pi(X); \alpha] = \frac{\sum_{i | [\hat{\pi}(X_i) \leq \hat{F}_\pi^{-1}(\alpha)]} \hat{\pi}(X_i)}{\sum_{i=1}^n \hat{\pi}(X_i)}$$

to as an independent line in the literature. Because If two variables Y and $\pi(X)$ are mutually independent, then $CC[Y, \pi(X); \alpha] = \frac{E[Y]P[\pi(X) \leq F_\pi^{-1}(\alpha)]}{E[Y]} = \alpha$.

The positive dependency of variables gives a convex concentration curve. Conversely, if the CC is convex, then the two variables are positively dependent. Nevertheless, the predictor $\pi_1(X_1)$ is more discriminatory than $\pi_2(X_2)$ for response Y if and only if the following inequality exist for all levels of $\alpha$.

$$CC[Y, \pi_1(X_1); \alpha] \leq CC[Y, \pi_2(X_2); \alpha] \quad (3)$$

In other words, CC curve of predictor $\pi_2$ is below the CC curve of predictor $\pi_1$. If the respective concentration or Lorenz curves of two predictor intersect, then ICC index is used rather than CC index to compare the discriminatory of two predictor (for more details, see Denuit et al., 2019).

Owing to Lorenz curve is a special case of concentration curve, in addition to the properties of the concentration curve, it has its own special properties. LC is derived by dividing the cumulative value of the variable by its expected value. That is related to the Gini's mean difference (GMD)[1] and Gini coefficient. The ratio of area is between 45-degree line (equality or identity line) and the LC over the total region under the line of equality, known as Gini coefficient. It can be shown that this region is equal to $2Cov[\pi(X), F_\pi(\pi(X))]$.

**Calculation of ABC Indicator**

The ABC indicator is given by Equation 4:

$$\begin{aligned} ABC[\pi(X)] &= \int_0^1 (CC[Y, \pi(X); \alpha] - LC[\pi(X), \pi\alpha]) \, d\alpha \\ &= \frac{1}{E[\pi(X)]} \int_0^1 (E[Y|[\Pi \leq \alpha]] - E[\pi(X)|[\Pi \leq \alpha]]) \, d\alpha \\ &= \frac{1}{E[\pi(X)]} \int_0^1 \int_0^\infty (P[\pi(X) \leq y, \Pi \leq \alpha] - P[Y \leq y, \Pi \leq \alpha]) \, dy \, d\alpha \\ &= \frac{1}{E[\pi(X)]} (cov[\pi(X), \Pi] - cov[Y, \Pi]) \end{aligned}$$

(4)

We use Equation 4 as a powerful model selection metric to decide which model is better than others.

---

[1]. As explained by Yitzhaki and Schechtman (2012), GMD has more than 14 alternative representations. The most convenient presentation of the GMD to be used is the covariance presentation, i.e. $E[|X_1 - X_2|] = 4Cov[X, F(X)]$.

## 3. Data and Estimation Results

### 3.1. The Descriptive Statistics of Research Variables

The raw data used in the research include both registered transaction information and macroeconomic variables. Table 1 shows the source and description of research variables. Data sample included 206,556 observed transaction data over the period from March 21, 2018 to February 19, 2020.

Table 1: Research variables and the relevant data sources

| Variables (Unit) | Description | Data Type | Source |
|---|---|---|---|
| Price (thousand IRR) | House price per square meter | | http://www.mrud.ir |
| Regional | Regional municipality | | |
| Region | Region (square meter) | | |
| Age | Building age (years) | | |
| Skeleton | Skeleton type: concrete, metal, brick or cement block, concrete and metal, skeletonless, clay, wooden | | |
| Dollar | Closed price of 1$ per IRR | Daily | https://www.tgju.org/ |
| Euro | Closed price of 1€ per IRR | Daily | |
| Emami coin | Closed price of 1Gold Emami coin per IRR | Daily | |
| TSE | Total price index of Tehran stock exchange | Daily | https://tse.ir/ |
| Land price (thousand IRR) | The average sale price of one square meter of land or residential building land | Quarterly | http://www.mrud.ir |
| Rent (IRR) | Average monthly rent plus 3% of the deposit payment on rent of 1 sq.m. | Quarterly | |
| CPI | Urban consumer price index (2016=100) | Monthly | https://www.amar.org.ir |
| Materials price | Building materials price index (2011=100) | Quarterly | http://www.mrud.ir |
| Age level | | | |
| Total price (thousand IRR) | Price * Region | | |

| En Date | Contract registration date | | |
|---|---|---|---|

Source: Research findings

The number of observed house transactions per municipality region of Tehran is plotted in Figure 1. In addition, the average price of transacted buildings, in each region is compared in Figure 2. Figure 2 illustrates how the brand of municipality affect the house price in Tehran. Figures 3 and 4 show that house details such as age and region vary across the municipality regional number. A possible explanation for this might be that some attributes of a house such as house region might be influenced by the brand of municipality.

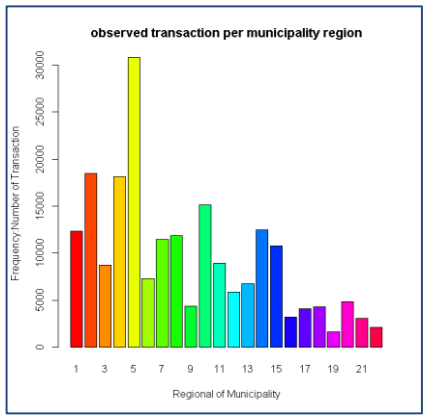

**Fig. 1: The number of observed transaction per region**

Source: Research findings

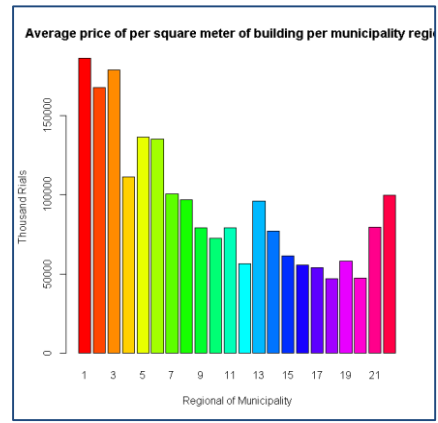

**Fig. 2: The comparison of average price of transacted building per region**

Source: Research findings

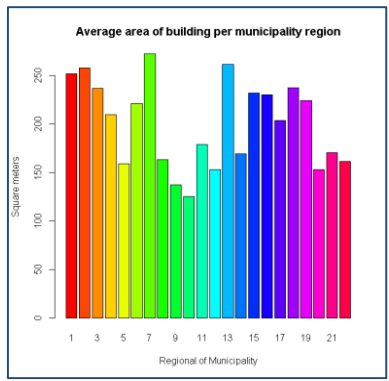

**Fig. 3: The comparison of average region of transacted building per region**

Source: Research findings

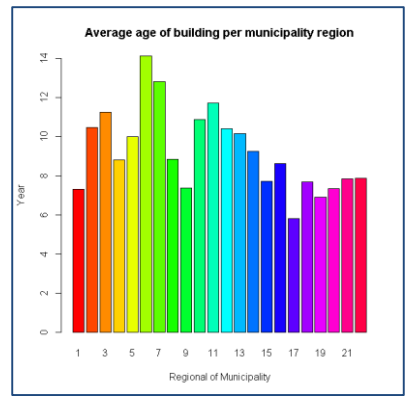

**Fig. 4: The comparison of average age of transacted building per region**

Source: Research findings

## 3.2. The Relevancy of the Selected Variables to the House Price

To identify the relevant house price factors, and detect the intensity and the direction of the effect of each factor on house prices, strength of association between house price and selected variables is measured (Table 2) by three different types of metrics: a) The Pearson correlation statistic used to explore the linear relationship. b) The Spearman's Rho rank based statistics used to assess the monotonic relationship (whether linear or not)[1]. c) The CC index to measure any type of dependency (whether linear or monotonic or non-monotonic)[2].

Following Denuit et al. (2019), to determine the most relevant variables to house price, the CC index of house price and the selected variable is computed in the first instance. Next variables are sorted and ranked by the absolute value of CC index.

Based on the statistical value of all three types of association metrics in Table 2, it is clear that house prices in Tehran are inversely related to regional municipality and building age. The relative importance of the variables in housing prices is presented in the third column of Table 2. According to the size of the CC index, the effect of land price variable on building price is more than other variables, so the land price is identified as the most "relevant" variable to the house prices in Tehran.

**Table 2: The values and comparison of the relevance of selected variables to house price**

| Selected variable | Concentration curve index | Relevancy rank | Pearson R2 | Spearman's Rho |
|---|---|---|---|---|
| Land price | 0.25840 | 1 | 0.1249 | 0.6862 |
| Rent | 0.25192 | 2 | 0.0786 | 0.4686 |
| Regional municipality | -0.18887 | 3 | -0.0922 | -0.5272 |
| CPI (urban consumer price index) | 0.16289 | 4 | 0.0769 | 0.4618 |
| Materials price index | 0.16289 | 5 | -0.0366 | -0.1699 |
| Stock price index (TSE) | 0.15952 | 6 | 0.1265 | 0.7186 |
| Gold price (Emami coin) | 0.13916 | 7 | 0.0726 | 0.4572 |
| Exchange rate $ | 0.12519 | 8 | 0.0580 | 0.3268 |
| Exchange rate € | 0.10962 | 9 | 0.0684 | 0.4038 |

---

[1]. That is equal to the Pearson correlation between the rank values of the two variables.

[2]. As mentioned in Section 2, the CC index is the copula of one variable (here house price) and the rank of another variable (such a house feature).

| | | | | |
|---|---|---|---|---|
| Building age (age level) | -0.06048 | 10 | -0.0277 | -0.1690 |
| Region | 0.05456 | 11 | 0.0657 | 0.3665 |
| Skeleton type | 0.01496 | 12 | -0.0104 | 0.3827 |

Source: Research findings

The concentration curves of house price with respect to each variable is plotted in Figure 5. It is apparent from Figure 1 that there is a significant difference in the concentration curve of house price determinant variables. The CC curve of some variables such as regional municipality and building age completely lie above the line 45-degree, owing to their negative effect on the house price. Some variables like land price and house rent lie below it due to the results of their positive effects. However, some variables might have a threshold effect on the house price. For example, the region of the house has threshold effect, therefore, its CC curve intersects the 45 degree line, and is located above and below the line.

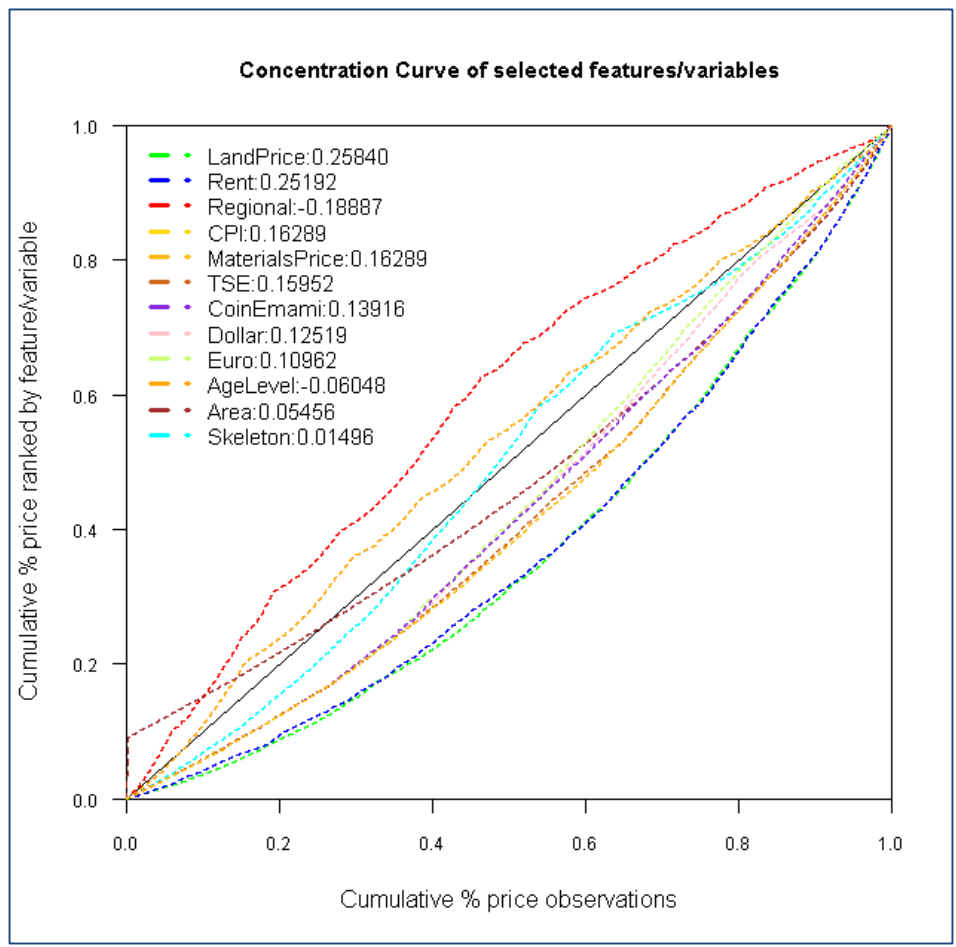

**Fig. 5: The comparison of concentration curve of house price to the selected variables**

Source: Research findings

### 3.3. Modeling the House Price Prediction

Any house price prediction models should consider the relevant house price determinant variables in the model. Therefore, here we setup the basic equation of house price prediction models based on the hedonic pricing equation as follows:

$$House\ price = f(Structural\ attributes,\\ Neighborhood\ attributes, Locational\ attributes,\\ Macroeconomic\ variables\ and\ Other\ factors) \quad (5)$$

The recorded and available information in the real estate transactions system have only allowed us to include the region of house, building age, and type of skeleton, as the structural details, and the number of the municipal location, as the locational attributes. However, to capture the effect of the costs of building construction on the housing transaction price, the variables such as building materials price index and land price are included in the prediction models. Furthermore, the house rent and the prices of foreign currencies (dollars and euros) and the stock market index and gold prices are included to capture effects of the optimal portfolio of investors of the competitive house market on the house price. Owing to the aims of the research, the above basic house price equation is estimated through the four different following approaches.

1. The GLM with two sub-families, i.e. gaussian GLM regression model and poisson GLM regression model;
2. RPART with two splitting rules of ANOVA and poisson models, i.e. "rpart.anova" and "rpart.poission";
3. RF regression models;
4. NN model with three types of hidden configurations (2,1), (3,1), and (3,2), i.e. "NN.21", "NN.31" and "NN.32" respectively.

As mentioned in Section 1, the research data sample is randomly divided into two sub-samples, i.e. "learn-sample" and "test-sample". These sub-samples cover 90% and 10% of all data samples used for estimation and prediction purposes, respectively[1]. To predict the house price, the following two-step process is conducted in each method approach:

1. Model construction and its estimation by using the learn-sample data;

2. Price forecast based on the estimated model by using test-sample data.

A summary of the estimation output of models is presented in the supplementary excel file. To compare the used models in modeling and forecasting the housing prices, Tables

---

[1]. It is noteworthy that in the estimation process of all models except neural network models, apart from building age, skeleton type, and regional municipality, other research variables are used in the natural logarithms form. However, for the neural network models, research variables are used in the normalization form which is normalized by max-min normalization technique.

3 and 4 summarize the descriptive statistics results of real prices with the fitted (or predicted) prices of each model based on both "learn-sample" and "test-sample" data sets, respectively. As indicated in Tables 3 and 4, the mean value of real price is less than its median value and, therefore, the distribution of observed house price is asymmetric with negative (left) skewness. However, according to the mean and median values, it seems that some models such as gaussian GLM and NN.32 give positive (right) skewness prediction. The minimum value and the 1st quantile value of fitted/predicted price is higher than the minimum value and the 1st quartile of real price, while the maximum value of fitted/predicted prices is less than the maximum value of real price.

The results in Tables 3 and 4 show that there is a significant difference between the statistical parameters of fitted/predicted prices and the statistical parameters of real prices in both "learn" and "test" samples.

**Table 3: The comparison of the descriptive statistics of actual and fitted value of price by "learn-sample"**

| | Statistical parameters | Min | 1st Quantile | Median | Mean | 3rd Quantile | Max |
|---|---|---|---|---|---|---|---|
| | Real price | 0.501 | 10.866 | 11.347 | 11.267 | 11.773 | 17.814 |
| Fitted value by: | Poisson GLM | 7.804 | 10.910 | 11.267 | 11.267 | 11.651 | 14.476 |
| | Gaussian GLM | 7.261 | 10.919 | 11.279 | 11.267 | 11.654 | 14.031 |
| | tree.anova | 10.440 | 10.980 | 11.270 | 11.270 | 11.620 | 12.050 |
| | tree.poisson | 10.440 | 10.980 | 11.270 | 11.270 | 11.620 | 12.050 |
| | RF | 4.138 | 10.909 | 11.308 | 11.266 | 11.662 | 13.818 |
| | NN.21 | 11.616 | 11.616 | 11.616 | 11.616 | 11.616 | 11.616 |
| | NN.31 | 11.605 | 11.605 | 11.605 | 11.605 | 11.605 | 11.605 |
| | NN.32 | 11.172 | 11.238 | 11.395 | 11.549 | 11.838 | 12.446 |

Source: Research findings

**Table 4: The comparison of the descriptive statistics of actual and predicted value of ln price by "test-sample"**

| | Statistical parameters | Min | 1st Quantile | Median | Mean | 3rd Quantile | Max |
|---|---|---|---|---|---|---|---|
| | Real price | 0.621 | 10.866 | 11.350 | 11.269 | 11.775 | 17.577 |
| Predi | Poisson GLM | 7.972 | 10.911 | 11.267 | 11.268 | 11.651 | 14.143 |

|  | | | | | | |
|---|---|---|---|---|---|---|
| Gaussian GLM | 7.498 | 10.919 | 11.280 | 11.268 | 11.654 | 13.766 |
| tree.anova | 10.440 | 10.980 | 11.270 | 11.270 | 11.620 | 12.050 |
| tree.poisson | 10.440 | 10.980 | 11.270 | 11.270 | 11.620 | 12.050 |
| RF | 4.494 | 10.910 | 11.309 | 11.267 | 11.659 | 13.843 |
| NN.21 | 11.616 | 11.616 | 11.616 | 11.616 | 11.616 | 11.616 |
| NN.31 | 11.605 | 11.605 | 11.605 | 11.605 | 11.605 | 11.605 |
| NN.32 | 11.172 | 11.237 | 11.391 | 11.548 | 11.836 | 12.445 |

Source: Research findings

### 3.4. Evaluation of the Models

To evaluate the performance of models, the accuracy of each price prediction model is measured with ABC criteria. Table 5 illustrates the estimated LC, CC, and ABC indices of the predicted price[1]. It also gives the rank of the models under study based on the forecasting accuracy. Furthermore, the concentration curves of the predicted price of all models in "test-sample" are illustrated in Figure 6. In Figure 6, the Lorenz curve of real price is plotted in black, but the concentration curves are drawn in color.

**Table 5: The comparison of the models' performance**

| Models | LC | CC | ABC | Accuracy rank |
|---|---|---|---|---|
| Poisson GLM | 0.391548 | 0.291207 | 0.100341 | 3 |
| Gaussian GLM | 0.391548 | 0.292435 | 0.099113 | 2 |
| rpart.anova | 0.391548 | 0.276273 | 0.115275 | 5 |
| rpart.poisson | 0.391548 | 0.276273 | 0.115275 | 6 |
| random.forest | 0.391548 | 0.325611 | 0.065937 | 1 |
| NN.21 | 0.391548 | -0.168691 | 0.560239 | 8 |
| NN.31 | 0.391548 | -0.098172 | 0.489720 | 7 |
| NN.32 | 0.391548 | 0.279327 | 0.112221 | 4 |

Source: Research findings

---

[1]. The estimated LC and CC indices of predicted price are used to calculate the ABC index.

As Table 5 shows, the prediction of some NN models such NN.21 and NN32, is inversely related to the real price data. Hence the estimated CC index for these models is a negative number. Therefore, the corresponding concentration curve of these models lies above the 45-degree line in the figure 6. Logically, the positive relationship between predicted prices and real prices is a necessity for choosing the suitable price prediction model. Therefore, two models of artificial neural network family with hidden configuration (2,1) and (3,2) are left out, and the more accurate prediction model is selected among other models based on ABC criteria. Both Table 5 and Figure 6 show that the accuracy of models in the prediction of house price, which is measured by ABC criteria, is not the same, and there is a significant difference between them. Interestingly, the lowest value of the calculated ABC index is related to the random forest model, as its related concentration curve is closer to the Lorenz curve. These results confirm that the selection of the random forest model as an eminently suitable model is a good choice for house price prediction.

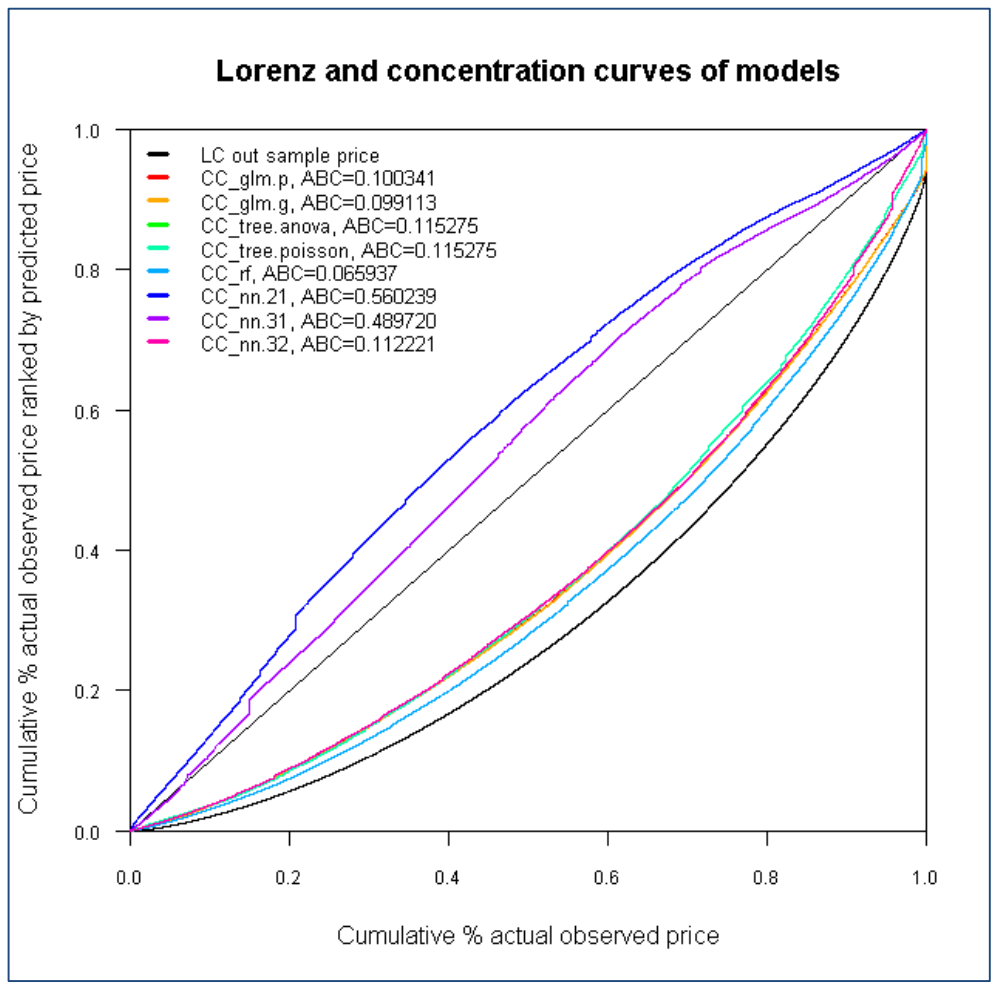

**Fig. 6: The comparison of the concentration curve of predicted price of all models**

Source: Research findings

## 4. Discussing Findings

Various models were found in the literature on the forecasting of house price. Furthermore, several different metrics were used to evaluate the performance of prediction models in the literature. The present study was designed to use a powerful model selection metric introduced by Denuit et al. (2019) to select the more accurate prediction model among several competitive house price prediction models. We found that accuracy of random forest method in the house price prediction was more than other models under study. Moreover, the GLM gives better prediction. It is somewhat surprising that random forest technique gives more accurate prediction of house price. This finding is consistent with that of Antipov and Pokryshevskaya (2012) who suggested the use of random forest models for tasks with missing values and categorical variables with many levels[1]. Additionally, this accords with the recent evidence found by Čeh et al. (2018) and Hong et al. (2020). This result may be explained by the fact that house price is affected by several categorical variables.

Another important finding of the present paper was the relevancy size of house price determining variables to the house price. Based on the CC index measure, the most positive relevancy is firstly with the land price and secondly with the house rent. As expected, the house price is negatively related to the regional municipality and the building age.

These findings are in line with those of previous studies (Sabbagh Kermani et al., 2010; Mohammadian Mosammam, 2015). Furthermore, the obtained results from Figure 5 showed that some variables such as region of house had threshold effects on the house price. This result may be explained by the fact that the purchasing power of households in purchasing the house was significantly unequal. Due to the low purchasing power of most residents of Tehran, the number of buyers of small houses is more than that of large houses.

These findings may help the policy makers to adopt appropriate tax policies in the housing market. Further work is required to establish the threshold effect of house price determinants.

## 5. Conclusion

This paper ranked several competitive house price prediction models, and highlighted the importance of a proper model selection in accurate house price prediction. Furthermore, we ranked the relevance of house price determinants variables to the house price in Tehran. The results of research were provided to select an eminently suitable model for housing price forecasting and to detect the main determinant of house price.

The current study used the ABC criterion to measure the accuracy of the predictors under study. It was not specifically designed to measure and compare the accuracy prediction of models with other performance metrics. As it did not addressed other price prediction models such as fuzzy logic and STSM models.

---

[1]. They used both coefficients of dispersion and MAPE indicators to compare the accuracy of different methods.

## 6. Appendix

Supplementary excel file

## 7. Research Highlights:
- Relevance of housing price determinant measured and ranked by concentration carve index.
- The more accurate price prediction model among several competitive considered models determined for Tehran housing market.
- The results of research provide to select an eminently suitable method/model for housing pricing/price forecasting and to determine the main determinant of house price.